\documentstyle[aps,12pt]{revtex}

\begin{document}
\author{Zhi-Jian Li$^1$\thanks{%
E-mail address: zjli5@yahoo.com}, Jiu-Qing. Liang$^1$, D. H. Kobe$^2$}
\title{Larmor precession and tunneling time of a relativistic neutral spinning
particle through an arbitrary potential barrier}
\address{$^1$Institute of Theoretical Physics, Shanxi University, Taiyuan Shanxi\\
030006, China\\
$^2$Department of Physics, University of North Texas, Denton, Texas\\
76203-5370, USA}
\maketitle

\begin{abstract}
The Larmor precession of a relativistic neutral spin-$\frac 12$ particle in
a uniform constant magnetic field confined to the region of a
one-dimensional arbitrary potential barrier is investigated. The spin
precession serves as a clock to measure the time spent by a quantum particle
traversing a potential barrier. With the help of general spin coherent state
it is explicitly shown that the precession time is equal to the dwell time.

PACS number(s): 03.65.Xp, 03.65.Ta
\end{abstract}

Over the years there have been many attempts\cite{1,2,3} to answer the old
and fundamental question ``how long does it take on average for an incident
particle to tunnel through a potential barrier?''. In the literature at
least three main approaches have been proposed to define and evaluate this
traversing time. First, one can study evolution of the wave packets through
the barrier and get the phase time\cite{4} which involves the phase
sensitivity of the tunneling amplitude to the energy of the incident
particle. Another definition of tunnel time is based on the determination of
a set of dynamic paths. The time spent in the different paths is averaged
over the set of the paths\cite{5}. The third approach makes use of a
physical clock to measure the time elapsed during the tunneling\cite{6,7,8,9}%
. The various approaches corresponding to different criteria have no a clear
consensus\cite{1,2,3}. Consequently, there has been existed much controversy
on the question of tunnel time. Recently a number of experiments\cite
{10,11,12} indicating superluminal transmission of photons through barriers
has renewed interest in the subject of tunneling time. Most of
investigations are focused on nonrelativistic tunneling and little work has
been done toward the study of relativistic tunneling time. The tunneling
time of the photons or the electromagnetic wave is usually investigated in
term of the mathematical identity of Helmholtz and Schr\"{o}dinger equations%
\cite{r1,r2} and the tunnelling time of Dirac electrons is calculated in the
framework of the Dirac equation in Refs.\cite{r3,r4}. In our earlier paper
we have reconsidered the Larmor precession of a neutral spinning particle in
a general spin coherent state as a clock to measure the tunneling time
through a one-dimensional rectangular barrier in the relativistic regime\cite
{13}. The reason that we choose the neutral particle instead of electron is
nothing but for the sake of simplicity, since the precession of spin in
magnetic field is the only effect to be considered for the Larmor clock
time. The potential barrier to a neutral particle can be simply a planar
film consisting of a medium characterized by a certain scattering potential
which has been well used in experiments of neutron resonators and
interferometers\cite{r8}. We now extend the previous investigation\cite{13}
to the case of an arbitrary potential barrier.

Larmor precession was first introduced long ago as a thought experiment
designed to measure the time associated with scattering events\cite{6}.
Subsequently the method was applied to measure the tunneling time of
particles penetrating barrier with a magnetic field confined to the barrier
region, causing the spin of particle to precess\cite{7}. The original scheme
in Ref.[7] considered only the rotation of the spin in the plane which is
perpendicular to the magnetic field. Later it was recognized that a particle
tunneling through a barrier in the magnetic field does not actually perform
a simple Larmor precession in a plane\cite{9}. The main effect of the
magnetic field is to align the spin along the field since the particle with
spin parallel to the magnetic field has lower energy and less decay rate in
barrier region than that of particle with spin antiparallel to the magnetic
field. In the present paper we use the spin coherent state to obtain an
equation of motion for the expectation value of spin operator in the
magnetic field. We show that the relativistic neutral spin-1/2 particles
perform a simple Larmor precession in three-dimensional space and that the
Larmor precession time equals the dwell time, which measures how long the
matter wave remains in the potential barrier regardless of whether the
particle is reflected or transmitted\cite{14}. For the special case of
symmetric potential barrier, the consistency of the dwell time, the
transmission time and the reflection time is obtained , which is in
agreement with the result for Schr\"{o}dinger particles\cite{1,9,r5}.

A relativistic neutral particle of spin $\frac 12$ with mass m and magnetic
moment $\mu $, moving in an external electromagnetic field denoted by the
field strength tensor $F_{\mu \nu }$, is described by a four-component
spinor wave function $\psi $ obeying the Dirac-Pauli equation\cite{15}

\begin{equation}
\lbrack \gamma ^{{\it \mu }}\frac{c\hbar }i\partial _{{\it \mu }}+mc^2+\frac %
12{\large \mu \sigma }^{{\it \mu \nu }}F_{{\it \mu \nu }}]{\normalsize \psi
=0,}
\end{equation}
where $c$ is the velocity of light in vacuum, $\gamma ^\mu =(\gamma ^0,{\bf %
\gamma })$ are Dirac matrices satisfying

\begin{equation}
\{\gamma ^\mu ,\gamma ^\nu \}=2g^{\mu \nu }
\end{equation}
with g$^{\mu \nu }$=diag$(1,-1,-1,-1)$, and

\begin{equation}
\sigma ^{{\it \mu \nu }}=\frac i2[\gamma ^{{\it \mu }},\gamma ^{{\it \nu }}].
\end{equation}
It can be shown that

\begin{equation}
\frac 12\sigma ^{{\it \mu \nu }}F_{{\it \mu \nu }}=i{\bf \alpha \cdot
E-\Sigma \cdot B}
\end{equation}
where {\bf E }and{\bf \ B }are the external electric and magnetic fields,
respectively, and $\alpha _i=\gamma ^0\gamma _i$, $\beta =\gamma ^0$ for
i=1,2,3. Here we make use of the Pauli representation

\begin{equation}
\beta =\left( 
\begin{array}{ll}
1 & 0 \\ 
0 & -1
\end{array}
\right) ,\ \alpha _i{\bf =}\left( 
\begin{array}{ll}
0 & \sigma _i \\ 
\sigma _i & 0
\end{array}
\right) ,\Sigma _i=\left( 
\begin{array}{ll}
\sigma _i & 0 \\ 
0 & \sigma _i
\end{array}
\right) .
\end{equation}
The spin operator is $S_i=\frac \hbar 2\Sigma _{i\text{ }}$and $\sigma _{i%
\text{ }}$are the Pauli spin matrices.

A incoming wave of relativistic neutral spin-$\frac 12$ particle polarized
in an arbitrary axis impinges on a finite range barrier potential $U(x)$
that extends from $a$ to $b$. A weak uniform constant magnetic field {\bf B ,%
} aligned along the z-direction and confined within the barrier region,
superimposes the barrier region (see Fig.1). The Hamiltonian is seen to be

\begin{eqnarray}
H_D &=&c\alpha _1p_x+\beta mc^2,\qquad \qquad x<a\text{ , }x>b \\
H_D &=&c\alpha _1p_x+\beta [(mc^2+U(x))-V\Sigma _3],\text{\qquad }a<x<b.
\end{eqnarray}
where $V=\frac \hbar 2\omega _L$ represents the spin-field interaction, $%
\omega _L=\frac{2\mu B}\hbar $ is the Larmor frequency, and $\hbar $ is
Planck's constant (divided by $2\pi $).

In the asymptotic regions $x<a$ and $x>b,$ the wave function satisfying the
stationary Dirac-Pauli equation

\begin{equation}
H_D\psi =E\psi ,
\end{equation}
is

\begin{eqnarray}
\psi _a &=&\frac 1{\sqrt{1+f_0^2}}\left( 
\begin{array}{l}
u_1 \\ 
u_2 \\ 
f_0u_2 \\ 
f_0u_1
\end{array}
\right) e^{\frac{ik_0x}\hbar }+\left( 
\begin{array}{l}
R_{U-V}\text{ }u_1 \\ 
R_{U+V}\text{ }u_2 \\ 
-f_0R_{U+V}\text{ }u_2 \\ 
-f_0R_{U-V}\text{ }u_1
\end{array}
\right) e^{-\frac{ik_0x}\hbar },\text{ \quad }x<a, \\
\psi _b &=&\left( 
\begin{array}{l}
T_{U-V}\text{ }u_1 \\ 
T_{U+V}\text{ }u_2 \\ 
f_0T_{U+V}\text{ }u_2 \\ 
f_0T_{U-V}\text{ }u_1
\end{array}
\right) e^{\frac{ik_0x}\hbar ,}\text{ \quad }x>b,
\end{eqnarray}
where 
\begin{eqnarray}
f_0 &=&\frac{ck_0}{mc^2+E}, \\
k_0 &=&\frac 1c\sqrt{E^2-(mc^2)^2}.
\end{eqnarray}
The quantities $T_{U\pm V}$ and $R_{U\pm V}$ denote the transmission and
reflection amplitudes, respectively, of an outgoing wave corresponding to
the total potential energy $U(x)\pm V$.

The incoming wave, i.e., the first term on the right hand side of Eq.(9), is
assumed to be a normalized spin coherent state which is an eigenstate of the
spin operator ${\bf \sigma \cdot n}$, where ${\bf n=(}\sin \theta \cos
\varphi ,$ $\sin \theta \sin \varphi ,$ cos$\theta {\bf )}$ denotes the
arbitrary unit vector with a polar angle $\theta $ and azimuthal angle $%
\varphi $\cite{16}. The two components of the spinor are

\begin{equation}
u_1=\cos \frac \theta 2e^{-i\varphi /2},\text{ \quad }u_2=\sin \frac \theta 2%
e^{i\varphi /2}.
\end{equation}
From the viewpoint of scattering, the outgoing wave consists of both a
reflected and a transmitted waves, which are separated from each other. The
outgoing wave must be normalized to unity since the incoming wave is
normalized to unity. The conservation of probability requires that the
coefficients $R_{U\pm V}$, $T_{U\pm V}$ satisfy the following relation

\begin{equation}
(1+f_0^2)(|T_{U\pm V}|^2+|R_{U\pm V}|^2)=1.
\end{equation}

For the case that V is small, i.e., the probing magnetic field is weak, $%
T_{U\pm V}$ and $R_{U\pm V}$ can be expanded as a power series of V to the
first order for the infinitesimal field approximation, such that

\begin{eqnarray}
T_{U\pm V} &=&|T_{U\pm V}|e^{i\alpha _{\pm }}\approx \left( |T_U|\pm V\frac{%
\partial |T_{U\pm V}|}{\partial V}\right) e^{i\left( \alpha _U\pm V\frac{%
\partial \alpha _{\pm }}{\partial V}\right) },  \nonumber \\
R_{U\pm V} &=&|R_{U\pm V}|e^{i\beta _{\pm }}\approx \left( |R_U|\pm V\frac{%
\partial |R_{U\pm V}|}{\partial V}\right) e^{i\left( \beta _U\pm V\frac{%
\partial \beta _{\pm }}{\partial V}\right) }.
\end{eqnarray}
(For the sake of brevity, we adopt the convention that the derivative $\frac %
\partial {\partial V}$ with respect to the auxiliary potential V is taken at
the zero field V=0, i.e., $\frac \partial {\partial V}|_{V=0}.$) The
transmitted wave and the reflected wave, $\psi _t$ and $\psi _r$
respectively, are

\begin{equation}
\psi _t=\left( 
\begin{array}{l}
T_{U-V}\text{ }u_1 \\ 
T_{U+V}\text{ }u_2 \\ 
f_0T_{U+V}\text{ }u_2 \\ 
f_0T_{U-V}\text{ }u_1
\end{array}
\right) ,\text{ \qquad }\psi _r=\left( 
\begin{array}{l}
R_{U-V}\text{ }u_1 \\ 
R_{U+V}\text{ }u_2 \\ 
-f_0R_{U+V}\text{ }u_2 \\ 
-f_0R_{U-V}\text{ }u_1
\end{array}
\right) .
\end{equation}
The expectation values of spin operator for the transmitted wave in the
infinitesimal field limit are

\begin{eqnarray}
\left\langle S_1\right\rangle _t &=&\frac \hbar 2(1+f_0^2)|T_U|^2\sin \theta
\cos (2V\frac{\partial \alpha _U}{\partial V}+\varphi ),  \nonumber \\
\left\langle S_2\right\rangle _t &=&\frac \hbar 2(1-f_0^2)|T_U|^2\sin \theta
\sin (2V\frac{\partial \alpha _U}{\partial V}+\varphi ), \\
\left\langle S_3\right\rangle _t &=&\frac \hbar 2(1-f_0^2)\left( |T_U|^2\cos
\theta -V\frac{\partial |T_U|^2}{\partial V}\right) .  \nonumber
\end{eqnarray}
The expectation values of spin operator for the reflected wave in the
infinitesimal field limit are

\begin{eqnarray}
\left\langle S_1\right\rangle _r &=&\frac \hbar 2(1+f_0^2)|R_U|^2\sin \theta
\cos (2V\frac{\partial \beta _U}{\partial V}+\varphi ),  \nonumber \\
\left\langle S_2\right\rangle _r &=&\frac \hbar 2(1-f_0^2)|R_U|^2\sin \theta
\sin (2V\frac{\partial \beta _U}{\partial V}+\varphi ), \\
\left\langle S_3\right\rangle _r &=&\frac \hbar 2(1-f_0^2)\left( |R_U|^2\cos
\theta -V\frac{\partial |R_U|^2}{\partial V}\right) .  \nonumber
\end{eqnarray}
Equations (17) and (18) show that the spin performs a Larmor precession
around the z-axis. To see the spin Larmor precession explicitly we may take
the sum of expectation values of spin in the reflected and transmitted
states. We then have

\begin{eqnarray}
\left\langle S_1\right\rangle &=&\frac \hbar 2\sin \theta \cos \left(
(1+f_0^2)|T_U|^22V\frac{\partial \alpha _U}{\partial V}+(1+f_0^2)|R_U|^22V%
\frac{\partial \beta _U}{\partial V}+\varphi \right) ,  \nonumber \\
\left\langle S_2\right\rangle &=&\frac \hbar 2\frac{(1-f_0^2)}{(1+f_0^2)}%
\sin \theta \sin \left( (1+f_0^2)|T_U|^22V\frac{\partial \alpha _U}{\partial
V}+(1+f_0^2)|R_U|^22V\frac{\partial \beta _U}{\partial V}+\varphi \right) ,
\\
\left\langle S_3\right\rangle &=&\frac \hbar 2\frac{(1-f_0^2)}{(1+f_0^2)}%
\cos \theta ,  \nonumber
\end{eqnarray}
where we have used Eq. (14) and the following power series with respect to
the small quantity V,

\begin{eqnarray}
&&(1+f_0^2)\left( |T_U|^2\cos (2V\frac{\partial \alpha _U}{\partial V}%
+\varphi )+|R_U|^2\cos (2V\frac{\partial \beta _U}{\partial V}+\varphi
)\right)   \nonumber \\
&\approx &\cos \left( (1+f_0^2)|T_U|^22V\frac{\partial \alpha _U}{\partial V}%
+(1+f_0^2)|R_U|^22V\frac{\partial \beta _U}{\partial V}+\varphi \right) 
\end{eqnarray}
and

\begin{eqnarray}
&&(1+f_0^2)\left( |T_U|^2\sin (2V\frac{\partial \alpha _U}{\partial V}%
+\varphi )+|R_U|^2\sin (2V\frac{\partial \beta _U}{\partial V}+\varphi
)\right)  \nonumber \\
&\approx &\sin \left( (1+f_0^2)|T_U|^22V\frac{\partial \alpha _U}{\partial V}%
+(1+f_0^2)|R_U|^22V\frac{\partial \beta _U}{\partial V}+\varphi \right) .
\end{eqnarray}
Equations (19) are formally the same as the Larmor precession equation of
spin ${\bf S}$ in a uniform constant magnetic field. To see this let us
consider a relativistic neutral spin-$\frac 12$ particle in a uniform
constant magnetic field B along the z-direction in the absence of potential
barrier. The Larmor precession is obtained by solving the Heisenberg equation

\begin{equation}
\frac d{dt}{\bf S(}t{\bf )=}\frac 1{i\hbar }[{\bf S(}t{\bf ),}H_s]
\end{equation}
with the spin Hamiltonian

\begin{equation}
H_s=-\frac 12\hbar \omega _L\beta \Sigma _3.
\end{equation}
If the initial wave function is given by the spin coherent state

\begin{equation}
\psi _i=\frac 1{\sqrt{1+f_0^2}}\left( 
\begin{array}{l}
u_1 \\ 
u_2 \\ 
f_0u_2 \\ 
f_0u_1
\end{array}
\right) ,
\end{equation}
the expectation values of the spin at time t are

\begin{eqnarray}
\left\langle S_1(t)\right\rangle &=&\frac \hbar 2\sin \theta \cos (-\omega
_Lt+\varphi ),  \nonumber \\
\left\langle S_2(t)\right\rangle &=&\frac \hbar 2\frac{1-f_0^2}{1+f_0^2}\sin
\theta \sin (-\omega _Lt+\varphi ),  \nonumber \\
\left\langle S_3(t)\right\rangle &=&\frac \hbar 2\frac{1-f_0^2}{1+f_0^2}\cos
\theta .
\end{eqnarray}

Comparing Eqs.(19) and (25), we see that for the infinitesimal magnetic
field the Larmor tunneling time $\tau _L$ is

\begin{equation}
\tau _L=(1+f_0^2)|T_U|^2\left( -\hbar \frac{\partial \alpha _U}{\partial V}%
\right) +(1+f_0^2)|R_U|^2\left( -\hbar \frac{\partial \beta _U}{\partial V}%
\right)
\end{equation}
which is just the average time scale over the transmission and reflection
channels defined as the dwell time $\tau _D$ \cite{1,17}, 
\begin{equation}
\tau _L=\tau _D.
\end{equation}
The transmitted time $\tau _T$ and the reflected time $\tau _R$ are
identified from Eq. (26) as

\begin{equation}
\tau _T=-\hbar \frac{\partial \alpha _U}{\partial V}\text{, \qquad }\tau
_R=-\hbar \frac{\partial \beta _U}{\partial V}\text{,}
\end{equation}
which are exactly in accordance with the Larmor time $\tau _T^y$ and $\tau
_R^y$ for the Schr\"{o}dinger particles introduced by B\"{u}tticker in Ref.
[9].

For the special case of symmetric potential barriers, i.e., $U(x)=U(-x)$,
the scattering phases $\alpha _U$ and $\beta _U$ satisfy the relation\cite
{17}

\begin{equation}
\alpha _U=\frac \pi 2+\beta _U.
\end{equation}
From Eq. (26) and the probability conservation Eq. (14) we obtain

\begin{equation}
\tau _L=\tau _D=\tau _T=\tau _R,
\end{equation}
which is the same result as obtained for the symmetric rectangular potential
barrier\cite{13}.

In Ref. [17] we have shown that the transmission time $\tau _T$ of the
relativistic neutral particles described by plane waves is given by

\begin{equation}
\tau _T=\frac{f_0}{c^2k}\frac{4dk\xi E(k^2-f_0^2\xi ^2)+\hbar (ck^2+E\xi
)(k^2+f_0^2\xi ^2)\sinh (\frac{4dk}\hbar )}{4f_0^2\xi ^2k^2+(k^2+f_0^2\xi
^2)^2\sinh ^2(\frac{2dk}\hbar )}
\end{equation}
for the symmetric rectangular potential barrier with the width 2d and height 
$U_0$, where

\begin{equation}
\xi \equiv \frac 1c(mc^2+U_0+E)\text{, \quad }k=\frac 1c\sqrt{%
(mc^2+U_0)^2-E^2}\text{.}
\end{equation}
We have demonstrated with numerical estimation that this transmission time
can be much smaller than the time for the particle to penetrate a constant
magnetic field without a barrier, which implies apparent superluminal
tunneling.

To summarize, we have presented a general proof that the dwell time equals
the Larmor precession time for a relativistic neutral spinning particle
penetrating an arbitrary potential barrier. We also extend the equality to
include the transmission time $\tau _T$ (i.e., Eq.(30)) explicitly for a
symmetric potential barrier.

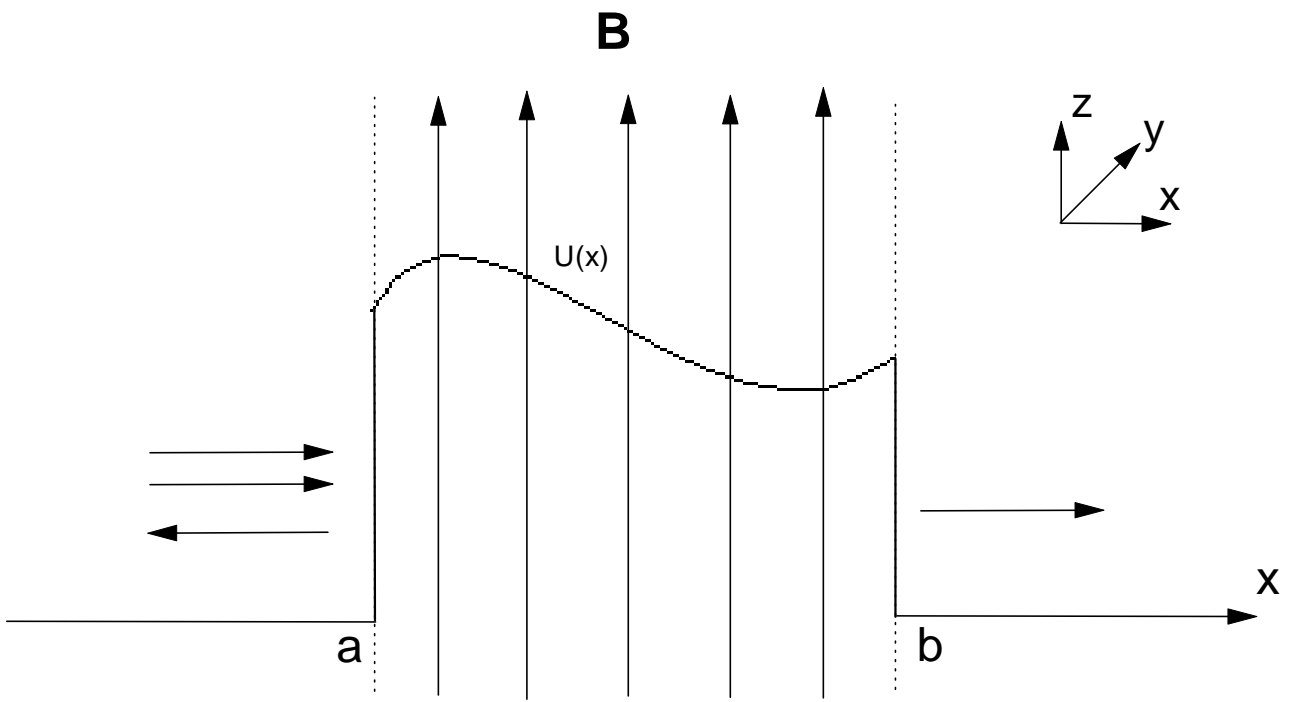
\end{document}